\documentclass[fleqn,usenatbib]{mnras}

\usepackage{newtxtext,newtxmath}

\usepackage[T1]{fontenc}
\usepackage{ae,aecompl}

\usepackage{graphicx}	
\usepackage{amsmath}	
\usepackage{amssymb}	

\newcommand{\be}{\begin{equation}}
\newcommand{\ee}{\end{equation}} 
\newcommand{\bea}{\begin{eqnarray}}
\newcommand{\eea}{\end{eqnarray}}
\newcommand{\PSD}{{\sc psd }}
\title[Discoseismic modes and turbulence]{On twin peak quasi-periodic oscillations resulting from
the interaction between discoseismic modes and turbulence in accretion discs
around black holes} 

\author[M. Ortega-Rodr\1guez et al.]{
M. Ortega-Rodr\1guez,\thanks{E-mail: manuel.ortega@ucr.ac.cr}  
H. Sol\1s-S\'anchez, 
L. \'Alvarez-Garc\1a,  
E. Dodero-Rojas
\\
Laboratorio de F\1sica Te\'orica 
\& 
Centro de Investigaciones Geof\1sicas, Universidad de Costa Rica, 11501-2060 San Jos\'e, Costa Rica
}

\date{Accepted 2019 December 11. Received 2019 December 11; in original form 2019 August 8} 

\pubyear{2019}

\begin{document}
\label{firstpage}
\pagerange{\pageref{firstpage}--\pageref{lastpage}}
\maketitle

\begin{abstract}
Given the peculiar and (in spite of many efforts) unexplained 
quasi-periodic oscillation (QPO) twin peak phenomena in accretion disc \PSD observations,
the present exploratory analytical article tries to inquire deeper
into the relationship
between discoseismic modes and the underlying driving turbulence 
in order to assess its importance. 
We employ a toy model in the form of a 
Gaussian white noise driven damped harmonic oscillator 
{\it with stochastic frequency}. 
This oscillator represents the discoseismic mode. 
(Stochastic {\it damping} was also considered, but interestingly found 
to be less relevant for the case at hand.) 
In the context of this model, 
we find that turbulence interacts with disc oscillations in 
interesting ways. 
In particular, the 
stochastic part in the oscillator frequency
behaves as a separate driving agent. This gives
rise to 3:2 twin peaks for some values of the physical parameters, which we find. 
We conclude with the suggestion that the study of turbulence 
be brought to the forefront of disc oscillation dynamics, 
as opposed to being a mere background feature. 
This change of perspective carries immediate observable consequences, 
such as considerably shifting the values of the
(discoseismic) oscillator frequencies. 
\end{abstract}

\begin{keywords}
accretion, accretion discs -- black hole physics -- X-rays: binaries
\end{keywords}



\section{INTRODUCTION}


After two decades, 
and in spite of active research,
the remarkable structure in the power spectra of
high frequency (40-450 Hz) quasi-periodic oscillations (HFQPO) in 
several X-ray binaries remains
an intriguing puzzle.
This structure often consists of 
power spectra 
twin peaks in a 3:2 frequency ratio 
(\citealt{Abramowicz};
see also 
\citealt{Remillard3};
\citealt{Homan}; 
\citealt{Belloni2};
for a full list, refer to 
\citealt{Ortega} and references therein).
 
An understanding of HFQPOs may allow us to
obtain important information about the 
corresponding black hole's 
mass and spin, and the behavior of inner-disc accretion flows.

The physics of HFQPOs is not completely understood.
Since the observed 3:2 frequency ratio suggests the presence of
non-linear physics, resonant models have been proposed
(see e.g.  
\citealt{Kluzniak};  
\citealt{Stuchlik}; 
a detailed discussion of models can be seen in 
\citealt{Torok2}). 

We wish to consider here a different line of inquiry,
which does not exclude non-linear considerations 
but rather could work in tandem with them.

The main objective of this paper
is to explore
the question of what happens when 
the MRI-generated turbulent background in an
accretion disc interacts with a
discoseismic fundamental g-mode, 
or any type of sufficiently stable oscillation
for that matter (we discuss below what has been researched so far).
For a review of discoseismology, see \citet{Wagoner1};
it is important to mention that while
discoseismic modes were reported to be present in hydrodynamic simulations (\citealt{Reynolds}), 
they have yet to be seen in MHD simulations. 
 


We implement this interaction in the form of a toy model.
When devising a toy model
for the disc's complicated dynamics,
our aim was to propose
the simplest mathematical expression
which includes the main physics: 
turbulence, mode oscillation, plus
the turbulence-induced frequency variability of this mode oscillation. 
(For completeness, damping variability was also included.)   
We aimed thus for a single differential equation: 
\be\label{main} 
   \ddot{x} + \alpha(t) \dot{x} + \omega^2(t) x = f(t) \, ,
\ee
where the driving term
$f(t)$, representing turbulence, is 
(within our toy framework) 
Gaussian white 
noise (see, however, below) with zero mean and 
\be\label{gaussian} 
   \langle f(t) f(t^\prime) \rangle = {\rm constant} \cdot \delta(t-t^\prime) 
\ee
(brackets standing for ensemble average), whereas 
$\omega(t)$, $\alpha(t)$, 
representing the mode frequency and damping coefficient, 
are randomly varying functions of time 
(in a way specified below).


 
At this point it is sensible to ask whether
the single equation (\ref{main}) is by itself enough to 
capture the necessary insights to understand this physical system. 

We believe that, when modeling the system, 
the mental picture one should have in mind is less 
a set of coupled oscillators (one for the mode, one for turbulence) 
exchanging energy back and forth, 
and more one of a mostly unidirectional transfer of energy in which
turbulence acts effectively as a reservoir whose energy
is fed by a (presumably MRI driven)   
cascading Kolmogorov process from larger to smaller scales. 

Furthermore, note that the effects of turbulence enter in two places in equation (\ref{main}): 
in the RHS, via $f(t)$, and in the LHS, via the variability in $\omega$ and $\alpha$, 
and although $f(t)$ is white, the LHS terms effectively couple 
the modes only to certain frequencies in the turbulence, as discussed below. 
(One might ask why the need of $f(t)$ in the first place, and the reason is that it 
helps with stability of the system; see e.g. \citealt{Frisch}.)
 



With these considerations, 
we will find that turbulence interacts with randomly-variable oscillations in 
interesting ways, and this  
gives a possible explanation for twin peak QPOs.  
We conclude that turbulence might be important in these dynamics
and suggest that its study be brought to the forefront, 
as opposed to being a mere background feature. 
This change of perspective brings immediate observable consequences, 
such as considerably shifting the values of the 
(discoseismic) oscillator frequencies. 

It is important to compare our approach with that of \citet{Vio}.  
Their toy model uses {\it two} coupled non-linear oscillator equations 
of {\it constant} (or non-stochastically variable) 
frequency with a stochastic source for one of them. 
Their results are interesting because they find that such a turbulent source does enhance
both oscillations, and that there is a window of opportunity
(too much or too little turbulence result in no enhancing).
As these findings are also results of the present paper, both papers taken 
together do make a strong case for the importance of turbulence in these 
astrophysical systems.  
The main difference between the articles is that we only assume a single oscillator, 
so the appearance of the second frequency in the spectrum is more intriguing
and meaningful. 

\section{THE MODEL}

\subsection{Parametrizing stochasticity}

\noindent
The approach starts from oscillator equation (\ref{main}), 
which is a generalization of the one developed by \citet[hereafter BFP]{Bourret} in
that it allows for a time dependent $\alpha(t)$.  
The rationale is that turbulence is as likely to modify $\alpha$
as to modify $\omega$. 

The stochasticity of $\omega$ is parametrized thus:
\be
      \omega^2(t) = \omega_0^2[1+\varepsilon m(t)] \ ,
\ee
where $\omega_0$ is a constant, $\varepsilon$ is (without loss of generality) a positive number and
$m(t)$ is a two-valued Markov process variable taking the values $\pm 1$ 
(central limit theorem considerations render this approach less toyish than it 
might seem at first sight). 
Concerning $m(t)$, its first important property is
\be\label{ensemble}
  \langle m(t) \rangle = 0 \, .   
\ee
Brackets stand for ensemble average.
The ensemble average of a quantity $W$ which depends on a set of two-valued variables $m_i$ 
[so that $W(m_i)$ stands for $W(m_1,m_2,...)$] is defined here as 
\be
   \langle W(m_i) \rangle \equiv \sum P(m_k) W(m_k) \, , 
\ee 
where the sum runs over all specific combinations of the $m_i$ set, and 
$P(m_k)$ is the probability of the specific combination of $m_k$  values. 

With this definition, equation (\ref{ensemble}) follows trivially.
For averages involving variables at different times, we will need to introduce 
the corresponding conditional probabilities: $(s, \nu > 0)$ 
\begin{align}
{\rm Prob}\{m(t+s) &= \pm1 \, | \, m(t) = \pm1\} = \tfrac{1}{2}(1 + e^{-\nu s}) \, , \\
{\rm Prob}\{m(t+s) &= \pm1 \, | \, m(t) = \mp1\} = \tfrac{1}{2}(1 - e^{-\nu s}) \, . 
\end{align}
These equations can be thought of as defining $\nu$.
We can now obtain the autocorrelation 
\be
  \langle m(t+s) \, m(t) \rangle = e^{-\nu |s|} 
\ee  
(the absolute value on $s$ allows it to take negative values). 

For the damping term, we proceed analogously:
\be
     \alpha(t) = \alpha_0[1+\delta n(t)] \ ,
\ee
where $\alpha_0$ is a constant, $\delta$ is a positive number and
$n(t)$ is a two-valued Markov process variable with the following properties: 
\be
 \langle n(t) \rangle = 0 \ , \quad  \langle n(t+s) \, n(t) \rangle = e^{-\bar{\nu} |s|} \, .
\ee

In a spirit of simplicity, 
we will work here the case in which 
$m(t)$ and $n(t)$ are uncorrelated, i.e. 
$\langle m(t) \, n(t^\prime)\rangle = 0$. 
 
\subsection{Physical meaning of the parameters}

We will now make explicit the correspondence `dictionary' relating the toy-model parameters
to their physical counterparts in the accretion disc system. 

\begin{enumerate}

\item $\omega_0$ refers to the mode frequency;
      this is not the same as the {\it observed} frequency, which is given
      by the effective frequency $\omega_{\rm eff}$, as explained below 
      (as usual, all frequency values in the present article refer to 
      those detected far away from the black hole). 

\item $\alpha_0/\omega_0 \approx 0.1$ is the inverse QPO quality factor $Q$; 
      note that this parameter and the previous one can be read off, 
      in an approximate fashion,   
      directly from a power spectral density ({\sc psd}) graph 
      (the numerical value $0.1$ is taken from typical observational data); 
      the reading is only approximate because  
      $\varepsilon$ can affect 
      the reading of $\alpha_0$, making it appear slightly larger than it is, 
      and turbulence effectively shifts the frequency, as explained below. 

\item the white noise $f(t)$ refers to the underlying turbulence; 
      its mathematical properies are defined by equation (\ref{gaussian}).

\item $\varepsilon$, defined as positive without loss of generality, 
      measures the variability of $\omega$; 
      we will place it in the range $0 < \varepsilon < 1$ because larger
      values of $\varepsilon$ are incompatible with observations 
      as they would force broad \PSD peaks with low values of $Q$. 

\item $\delta$, also defined positive, 
      measures the variability of $\alpha$; we also set it 
      in the range $0 < \delta < 1$ because its physical origin is presumably 
      the same as that of $\varepsilon$.

\item $\nu$ and $\bar{\nu}$ measure the time correlations 
      of the time-changing variables (frequency and damping coefficient, 
      respectively); scaling Kolmogorov considerations 
      (\citealt{Kolmogorov}; \citealt{Cho}) indicate
      that the coupling between oscillations such as fundamental discoseismic g-modes 
      and turbulence is  
      strong only when $\nu$, $\bar{\nu} \sim \omega_0$ or slightly smaller.  

\end{enumerate}

To elaborate on the last point, Kolmogorov scaling considerations assert 
that turbulent eddy frequencies ($\omega_{\rm tur}$) and length scales ($\lambda$) satisfy $\omega_{\rm tur} \propto \lambda^{-2/3}$ under broad conditions. 
The constant of proportionality is $u_* L_*^{-1/3}$, where $u_*$ and $L_*$ are the typical 
speed and length scales of the turbulent physical system. 
For thin disks, $u_* \sim h \Omega$ and $L_* \sim h$, where $h$ is the typical 
disc thickness and
$\Omega$ stands for the typical 
angular frequency. The typical angular frequency is also (approximately) the g-mode frequency,
i.e. $\Omega \approx \omega_0$.
This all means that setting $\lambda =$ size of fundamental g-mode $\sim \sqrt{hR}$, 
where $R$ is the typical radial scale, one has $\omega_{\rm tur} 
\sim  \omega_0 \, (h/R)^{1/3}$, which places 
the frequency of g-mode-sized turbulent eddies somewhat below $\omega_0$
(for $h/R = 0.1$, $\omega_{\rm tur} 
\approx 0.5 \, \omega_0$). These are the eddies that have a greater effect on 
discoseismic g-modes.

\section{RESULTS AND DISCUSSION}

\subsection{Frequency shift}

We then follow through the analytical method devised by BFP,
expanding it to include variable damping.
Since the approach is straightforward 
and grounded in well established mathematics, 
we merely give the results, 
leaving details for Appendix A.

We first look at what happens to the {\it observed} frequency after the effects
of stochasticity have been introduced. This effective  
frequency is given by  
\be
    \omega^2_{\rm eff} \equiv 
    \frac{\langle\dot{x}^2\rangle_{\rm eq}}{\langle x^2 \rangle_{\rm eq}} \, ,
\ee
where brackets indicate ensemble averages and `eq' indicates
that the system has been allowed to reach stationarity (and we therefore 
assume that it does).

For the special case of stochastic frequency and constant damping ($\delta = 0$),
we obtain
\be\label{omegaeff}
\omega_{\rm eff}^2 = \omega_0^2-\frac{2 \varepsilon^2 \omega_0^4 (2 \alpha_0+\nu)}{(\alpha_0+\nu)
   \left(2 \alpha_0 \nu+\nu^2+4 \omega_0^2\right)} \, , 
\ee
which reproduces equation (3.25) in BFP. 

In order of magnitude,
taking $\nu \sim \omega_0$ and $\alpha_0 \sim 0.1 \, \omega_0$, one obtains the
following: 
\be\label{omegaeffoom}
\omega_{\rm eff}^2 = \omega_0^2 \, [1 + {\cal{O}}(\varepsilon^2)] \, .  
\ee
This is an important result since 
$\omega_{\rm eff}$ is the frequency that detectors measure,  
and it can differ substantially from the theoretical one, $\omega_0$ 
(calculated, for example, from
discoseismology), given that $\varepsilon$ is not much smaller than 1. 
(When substituting the relevant numbers, one finds that the shift in 
frequency ranges from a few percent to $\approx$ 40\%.)  

We now turn to the
special case of constant frequency ($\varepsilon = 0$) 
and stochastic damping ($\delta \neq 0)$, for which one obtains
\be\label{omegaeff2}
\omega_{\rm eff}^2 = \omega_0^2-\frac{2 \alpha_0^2 \delta^2 \bar{\nu} \omega_0^2}{\bar{\nu} \left[2
   \alpha_0^2 \left(\delta^2+1\right)+3 \alpha_0 \bar{\nu}+\bar{\nu}^2\right]+4
   \omega_0^2 (\alpha_0+\bar{\nu})} \, .
\ee
In order of magnitude, we have 
\be\label{omegaeffoom2}
\omega_{\rm eff} = \omega_0^2 \, [1 + {\cal{O}}\{\delta^2(\alpha_0/\omega_0)^2\}] \, .
\ee
In this way, 
the variability of the damping coefficient
is less important, by two orders of magnitude, than the 
variability of the frequency. 
This points to a scenario in which the physics of stochasticity 
is contained already when considering just variable frequency. 
For this reason, from now on (unless otherwise indicated) we consider 
the $\delta=0$ case only. 

\subsection{Stationarity and stability}

As BFP correctly point out, procedures such as the one carried out in the present paper  
{\it assume} that there exist stationary solutions.
With this assumption one can obtain results like this one: ($\delta = 0$ case)
\be\label{x2} 
   \langle x^2 \rangle_{\rm eq} = \frac{S}{2\omega_0^2\alpha_{\rm eff}} \, ,
\ee
where
\be\label{alphaeff} 
   \alpha_{\rm eff} = \alpha_0 - \frac{\varepsilon^2\omega_0^2(\nu+2\alpha_0)^2}{(\nu+\alpha_0)
                                    [4\omega_0^2+\nu(\nu+2\alpha_0)]}
\ee 
(see Appendix A for details).

Self-consistency must be checked though. 
This result for $\langle x^2 \rangle$ does not make physical sense for some values of the 
parameters; in an obvious fashion, for those which render $\alpha_{\rm eff}$ negative,
and it is not clear that all solutions with positive $\alpha_{\rm eff}$ are stable.  
What one needs is a proof of stability. 
BFP actually prove that all solutions for which $\alpha_{\rm eff}$ 
is positive are indeed stable (see Appendix B for a discussion). 

We see then that for large enough values of $\varepsilon$ 
the system becomes unstable and therefore non-stationary. 
This happens for the critical value $\varepsilon_c$ (for which $\alpha_{\rm eff} = 0$) such that 
\be\label{epscrit}
   \varepsilon_{\rm c}^2 =  
     \frac{\alpha_0 (\alpha_0+\nu)[4\omega_0^2+\nu(\nu+2\alpha_0)]}
       {\omega_0^2(\nu+2\alpha_0)^2} \, .
\ee 

\subsection{Twin peaks}

There are two peaks in the \PSD of the stationary 
solution of equation (\ref{main}) for some values 
of the parameters. The way to obtain the formula for the \PSD is explicated in Appendix A, 
and the results are given by equations (\ref{Laplace2Fourier}) and (\ref{short}). 
Peak positions are obtained analytically by calculating the complex zeros of
the function $B$ defined in equation (\ref{denominator}).  

Furthermore, 
while peak frequencies can be in any ratio, 
for some values of the parameters this will be a 3:2 ratio. 
Fig.~1 shows the two peaks in a 3:2 ratio for values
of the parameters in the ranges discussed above. 

\begin{figure}
	\includegraphics[width=\columnwidth]{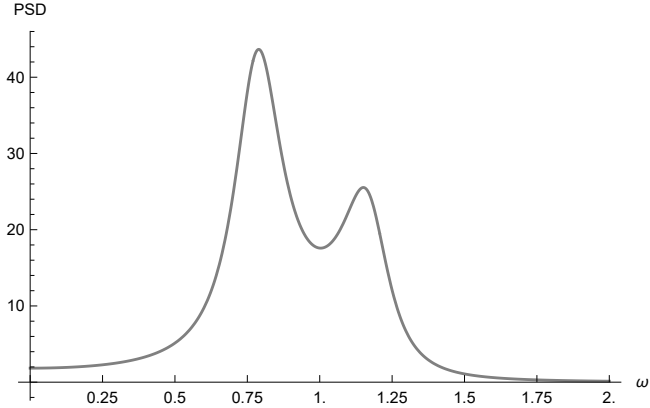}
    \caption{Power spectral density (arbitrary units) as a function of the frequency $\omega$ 
             (in units of $\omega_0$) 
             corresponding to the stationary stable solution of equation (\ref{main}) for the 
             following choice of parameters: 
             $\varepsilon = 0.4$, $\alpha_0 = \nu = 0.1 \, \omega_0$, $\delta = 0$. 
             The peak on the right arises purely as a result of the 
             stochasticity of the system.}
\end{figure}

As we mentioned, one way to understand 
the physical origin of the higher-frequency peak (the one not
arising from $\omega_0$) is by noting that the term
$\varepsilon m(t) \omega_0^2 x$ on the LHS of equation (\ref{main}) has, 
by virtue of its stochasticity, a behavior 
which is so different from the other one ($\omega_0^2 x$) 
that it behaves as a separate driving agent from $f(t)$.
Of course one needs the presence of the (non-stochastic) oscillator 
on the LHS of equation (\ref{main}) for this high-frequency peak to appear, no less 
than for the low-frequency one. 
Favoring this separate-agent interpretation is the fact that two-peakedness 
disappears for low enough $\varepsilon$, in which case
the stochastic term is `assimilated' by the dominant
$\omega_0^2 x$ term. 

The question might arise of how one knows that it is the {\it lower} 
(and not the upper) 
frequency peak the one
that corresponds to the $\omega_0$ term in equation (\ref{main}). The answer is that, as one 
computes the \PSD graph by scanning all possible values of the variables $\varepsilon$, $\nu$ and $\alpha_0$, 
one can see the appearance  
of the higher frequency peak branching off to the right side, 
and always from the bottom part,  
of the main one
(which is always present). By such a parameter scanning one can also appreciate 
that the lower frequency peak is always larger in amplitude than the 
higher frequency one; this makes sense as the lower frequency peak corresponds to the main $\omega_0$ oscillator 
in equation (\ref{main}). 

Fig.~2 shows the possible values of the parameters for which the frequencies of the \PSD peaks are 
in a 3:2 ratio.  
The vertical axis is $\varepsilon$; 
the horizontal axis is $\nu/\omega_0$; 
$\alpha_0$ is set to $0.1 \, \omega_0$ but the results are rather insensitive to its value.
We are interested in solutions lying in the shaded region.
All points which are above the solid straight line 
correspond to two-peak solutions
(as opposed to just one-peak), and
all points below the solid non-straight curve, 
which is given by $\varepsilon = \varepsilon_{\rm c}$ of 
equation (\ref{epscrit}), correspond to {\it stationary} solutions. 
Thus, the shaded region corresponds to stationary, two-peaked solutions. 
In order to be consistent with the discussion in item 2.2 (vi), we have left out as well 
the leftmost part of the region between the solid lines, say $\nu/\omega_0 < 0.1$; turbulence is not
an effective agent there. 

\begin{figure}
	\includegraphics[width=\columnwidth]{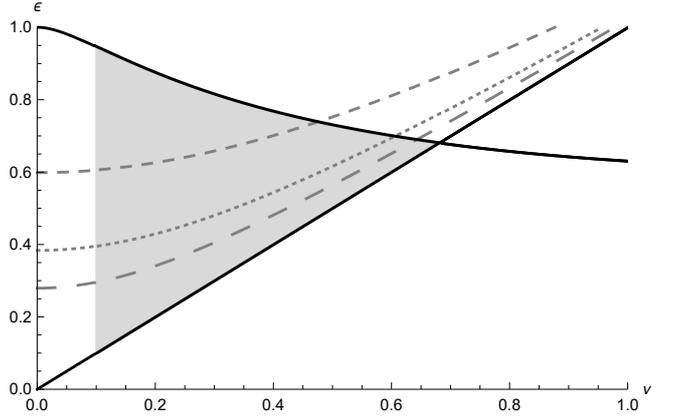}
    \caption{Values of $\varepsilon$ and $\nu$ (in units of $\omega_0$) 
             that produce peaks in a 3:2 frequency ratio (dotted line). 
             Also shown are the cases of 2:1 ratio (short dash) and 4:3 ratio (long dash). 
             The shaded region corresponds to stationary, two-peaked solutions with values of $\nu$ that
             allow mode-turbulence coupling. The region with low values of $\nu$ is excluded 
             in agreement with the qualitative discussion of item 2.2 (vi).}  
\end{figure}

The solutions in Fig.~2 corresponding to peaks in a 3:2 frequency ratio are given 
by the dotted curve (between the two dashed curves).  Also shown are solutions
corresponding to other ratios. 
It is important to emphasize that 
being in the shaded region is only 
a necessary condition for the existence of the twin peaks and
by all means not a sufficient one.
The peaks still need a mechanism to grow and become visible. 
This mechanism 
might be given, for example, by a non-linear resonance,
which is more effective for 
lower values of $a$, $b$ in an $a:b$ frequency ratio 
(see the references in the first section of this article). 

The above considerations give the 3:2 frequency ratio an optimal condition  
of having low integers while at the same time 
being compatible with larger values of $\nu$ (which imply a stronger coupling
between modes and turbulence), and being not too close to the 
bold straight line in Fig.~2 (so as to allow a clear valley between the peaks).   
 
Our considerations allow us to make the following prediction. 
As better X-ray timing observations become available, 
the lower frequency QPO will show a larger amplitude than the other twin.
Furthermore, the higher frequency QPO is the one that will intermittently 
appear and disappear, as it depends on favorable conditions of the variable turbulent 
dynamics. 

Future work could attempt to explore the physical system using an  
equation or set of equations having more structure than equation (\ref{main}), 
for example one based on hydrodynamical considerations. Numerical work can 
also complement this line of research.

\section*{Acknowledgements}

This work was supported in part by grant B6148 
of Vicerrector\'{\i}a de Investigaci\'on,
Universidad de Costa Rica. 
We are grateful to W\l{}odzimierz Klu\'zniak, Robert V. Wagoner, 
and Alexander Silbergleit 
for helpful comments. 








\appendix

\section{Derivation of formulas for the 
effective frequency and the PSD}

We follow, and generalize, the solution of BFP to  
equation (\ref{main}) under the conditions of stationarity of the 
solutions. 

The starting point is recasting equation (\ref{main}) in matrix
form:
\be\label{matrixform}
   \frac{dX}{dt} = M(t) \, X + F(t) \, ,
\ee
where
\be
    X \equiv 
\begin{pmatrix} 
x \\ 
\dot{x}   
\end{pmatrix} \, , \, \ 
M(t) \equiv 
\begin{pmatrix} 
0 & 1 \\ 
-\omega^2(t) & -\alpha(t)   
\end{pmatrix} \, , \, \ 
    F(t) \equiv 
\begin{pmatrix} 
0 \\ 
f(t)  
\end{pmatrix}  
\ee
(all variables were defined in the main
text).
Unlike BFP, we allow $\alpha(t)$ to depend on time. 

The quantities of interest take the form:
\be
  \Gamma_{ij}(t_1,t_2) \equiv \langle X_i(t_1) \, X_j(t_2) \rangle \, ,
\ee
and we note that in stationary regime  
$\Gamma_{ij}(t_1,t_2)$ by definition will depend only on the
difference $t_2-t_1 \equiv s$.
Thus, from now on, we 
will write $\Gamma_{ij}(s)$.

We are interested in particular in
$\Gamma_{11}(0)$ and $\Gamma_{22}(0)$ (section 3.1) and in 
the Laplace transform of $\Gamma_{11}(s)$ 
(section 3.3). 

We will use the Green function of equation (\ref{matrixform}), defined as the
matrix which satisfies
\be
  \frac{dG(t_1,t_2)}{dt_1} = M(t_1) \, G(t_1,t_2) \, ,  \quad G(t_1,t_1) = I \, ,
\ee
where $I$ is the $2 \times 2$ identity matrix. 
The function $\Gamma_{ij}(s)$ can now be expressed in terms of $G(t_1,t_2)$:
\be\label{Gamma}
\Gamma_{ij}(s) = S \int_0^t dt^\prime \langle G_{i2}(t,t^\prime) \, G_{j2}(t+s,t^\prime)\rangle \, ,
\ee
where $S$ is the constant that appears in equation  (\ref{gaussian}), and
the limit $t \rightarrow \infty$ is implied 
so as to ensure 
that the transient terms have vanished. 
Equation (\ref{Gamma}) can be put in a cleaner form noting that $G(t_1,t_2)$, being stationary, 
is invariant under time shifts. Subtracting then $t^\prime$ from all time arguments and introducing 
the time variable $T \equiv t-t^\prime$, one obtains 
\be\label{Gamma2}
\Gamma_{ij}(s) = S \int_0^\infty dT \langle G_{i2}(T,0) \, G_{j2}(T+s,0)\rangle \, .
\ee
Using this notation, we follow BFP and express everything in 
terms of the `vector' quantity $Z(t)$:
\be
  \frac{dZ(t)}{dt} = [L + m(t) L^\prime + n(t) L^{\prime\prime}] \, Z(t) \, ,
   \quad Z(0) = 
     \begin{pmatrix} 
       0 \\
       1 \\
       0 
     \end{pmatrix} \, ,
\ee
where 
\be Z(t) \equiv 
     \begin{pmatrix} 
       [G_{12}(t,0)]^2 \\
       [G_{22}(t,0)]^2 \\
       G_{12}(t,0) \, G_{22}(t,0)
     \end{pmatrix} \, ,
\ee
\be  L \equiv 
     \begin{pmatrix} 
       0 & 0 & 2 \\
       0 & -2 \alpha_0 & -2 \omega_0^2 \\
       -\omega_0^2 & 1 & -\alpha_0  
     \end{pmatrix} \, ,
\ee
\be  L^\prime \equiv -\varepsilon \omega_0^2   
     \begin{pmatrix} 
       0 & 0 & 0 \\
       0 & 0 & 2 \\
       1 & 0 & 0  
     \end{pmatrix} \, , \quad 
     L^{\prime\prime} \equiv -\delta \alpha_0    
     \begin{pmatrix} 
       0 & 0 & 0 \\
       0 & 2 & 0 \\
       0 & 0 & 1  
     \end{pmatrix} \, .
\ee
(We note in passing that there is a typographical error 
in the central term of $L$ in the first matrix of equation (3.10) in BFP;  
also note the difference in notation between their paper and ours.) 

We solve the differential equation in Laplace space.
The solution is
\be\label{M} 
   \langle \tilde{Z}(p) \rangle_i
   =
   [p-L-L^\prime(p+\nu-L)^{-1}L^\prime
   -L^{\prime\prime}(p+\bar{\nu}-L)^{-1}L^{\prime\prime}]^{-1}_{\phantom{i}i2} 
\ee
(recall that brackets refer to ensemble averaging), where 
\be
  \tilde{Z}(p) \equiv \int_0^\infty Z(t) \, e^{-pt} dt \, . 
\ee
We can finally write:
\be
   \langle x^2 \rangle_{\rm eq} = \Gamma_{11}(0) = S\int_0^\infty 
       \langle Z_1(t) \rangle \, dt = S  \, \langle \tilde{Z}_1(0) \rangle \, , 
\ee
\be 
   \langle \dot{x}^2 \rangle_{\rm eq} = \Gamma_{22}(0) = S\int_0^\infty 
       \langle Z_2(t) \rangle \, dt = S  \, \langle \tilde{Z}_2(0) \rangle \, ,  
\ee
which give us straightforwardly equations 
(\ref{omegaeff}) and (\ref{omegaeff2}).

The other quantity of interest is  
the Laplace transform of $\Gamma_{11}(s)$.
To obtain it, one starts form equation (\ref{Gamma2}) and 
(using the properties of the Green function) 
reexpresses it thus: 
\be\label{reexpress}
  \Gamma_{11}(s) = S \int_0^\infty [ \langle G_{11}(t+s,t) \, Z_1(t) \rangle + \langle G_{12}(t+s,t) \, Z_3(t) \rangle ] \, dt \, .
\ee
Working once more in Laplace space, and following a procedure analogous 
to the one of the preceding paragraphs, one obtains
\be\label{short}
\tilde{\Gamma}_{11}(p) = \frac{A}{B} (C+D) \ , 
\ee
where
\be 
 A(\alpha,\nu,\omega_0,\varepsilon) \equiv \frac{S}{2\omega_0^2} 
   \left( \alpha - 
     \frac{\varepsilon^2\omega_0^2(\nu+2\alpha)^2}{(\nu+\alpha)[4\omega_0^2+\nu(\nu+2\alpha)]} \right)^{-1} \, , 
\ee
\begin{multline}\label{denominator}
 B(p,\alpha,\nu,\omega_0,\varepsilon) \equiv
  p^4+2(\nu+\alpha)p^3+(\nu^2+3\alpha\nu+\alpha^2+2\omega_0^2)p^2  \\ 
 +(\nu+\alpha)(\nu\alpha+2\omega_0^2)p+\omega_0^4(1-\varepsilon^2) 
  +\nu(\nu+\alpha)\omega_0^2 \, ,
\end{multline}
\be 
 C(p,\alpha,\nu,\omega_0) \equiv  
  (p+\alpha)(p+\nu+\alpha)(p+\nu)+\omega_0^2(p+\alpha) \, , 
\ee
\be 
 D(p,\alpha,\nu,\omega_0,\varepsilon) \equiv  
  \frac{\varepsilon^2\omega_0^4(\nu+2\alpha)(2p+2\alpha+3\nu)}{(\nu+\alpha)(\nu^2+2\alpha\nu+4\omega_0^2)} 
  \, .
\ee
We have worked here the $\delta=0$ case, and thus $\alpha$ actually stands 
for $\alpha_0$ in the last four equations. 
Solving the $\varepsilon=0$, $\delta \neq 0$ case is straightforward 
but those results are astrophysically less relevant
as they have no measurable
consequences, as explained in the main text. (This means
that all which came after equation (\ref{reexpress}) is actually a 
restatement of BFP, and is included here for completeness.)

The above formalism allows us to express the {\sc psd} 
in terms of the Laplace transform of $\Gamma_{11}$. 
The {\sc psd} is the square modulus of $F(\omega)$, the Fourier transform  of
$x(t)$, i.e.  
\be\label{Laplace2Fourier}
  {\textsc{psd}} \equiv |F(\omega)|^2 = \int_{-\infty}^{+\infty} e^{i\omega s} 
     \langle x(t) \, x(t+s) \rangle_{\rm eq} \, ds 
     = 2 \, \tilde{\Gamma}_{11}(-i\omega) \, . 
\ee

Note that peak location is especially sensitive to the $B$ function as it appears 
in the denominator in equation (\ref{short}). 

\section{Proof of Stability}

It is important to prove that the allegedly stationary solutions are indeed stable. 

Rather than repeating the proof of stability for the stationary solutions 
of the $\varepsilon \neq 0$, $\delta=0$ case, which is described in detail in BFP, section 4,  
we offer here the 
proof for the $\varepsilon=0$, $\delta \neq 0$ case, which follows the same argument.
Future work includes proving the general case for which 
$\varepsilon \neq 0$ and $\delta \neq 0$. 

For the $\varepsilon=0$, $\delta \neq 0$ case, (\ref{x2}) holds but the analogous to  
(\ref{alphaeff}) is now
\be
   \alpha_{\rm eff} = \alpha_0 - 
   \frac{2\alpha_0^2\delta^2(2\alpha_0\bar{\nu}+\bar{\nu}^2
   +2\omega_0^2)}{\bar{\nu} [2\alpha_0^2(1+\delta^2) + 3\alpha_0\bar{\nu}+\bar{\nu}^2] + 4(\alpha_0+\bar{\nu})\omega_0^2} \, ,
\ee 
and the critical value of $\delta$ (at which $\alpha_{\rm eff}$ changes sign) is given by
\be\label{deltacrit}
   \delta_{\rm c}^2 = \frac{(\alpha_0+\bar{\nu})(2\alpha_0\bar{\nu}+\bar{\nu}^2
         +4\omega_0^2)}{2\alpha_0\bar{\nu}(\alpha_0+\bar{\nu})+4\alpha_0\omega_0^2}  \, .
\ee 

The proof of stability for $\delta < \delta_c$ is as follows. 
The starting point is the fact that 
the necessary and sufficient condition for stability is that 
the matrix that appears in (\ref{M}), with $\varepsilon = 0$, i.e. 
\be\label{sing} 
    [p-L-L^{\prime\prime}(p+\bar{\nu}-L)^{-1}L^{\prime\prime}]^{-1} \, , 
\ee 
does not have any singularities for Re($p$) > 0 
(otherwise the inversion of the Laplace transform meets divergences). 
The singularities occur for those values of $p$ which 
make the determinant of this matrix vanish. 
This determinantal equation takes the explicit form
\be\label{determinantal} 
H(p) - \alpha_0^{2} \delta^{2} J(p) + \delta^{4} K(p) = 0 \, , 
\ee 
where 
\begin{multline} 
H(p) = (\alpha_0 + p) (\alpha_0 +p+\bar{\nu}) \left(2 \alpha_0 p+ p^2 +4 \omega_0^2 \right) \\ 
  \times \, \left[(p+ \bar{\nu}) (2 \alpha_0 +p+ \bar{\nu})+4 \omega_0^2 \right] \, ,
\end{multline}
\begin{multline} 
J(p) = p (p+\bar{\nu}) \left[8 \alpha_0^2 +6 \alpha_0 (2 p+\bar{\nu})+5 p (p+\bar{\nu})\right] \\ 
   + 8 \omega_0^2 \left[\bar{\nu} (\alpha_0 +p)+p (2 \alpha_0+p)+ \bar{\nu}^2 \right]+16 \omega_0^4  \, , 
\end{multline} 
\be 
K(p) = 4 \alpha_0^4 p (p + \bar{\nu}) \, .
\ee 
 
The argument then is this: Firstly, prove that
there are no singularities with Re($p$) > 0 for 
sufficiently small $\delta$. Secondly, find the value of $\delta$ 
where the system becomes unstable by setting $p=0$ in 
equation (\ref{determinantal}). 
We do these steps in turn. 

First, for the no singularities part, proceed by {\it reductio ad absurdum}. 
Assume there is a solution of equation (\ref{determinantal}) such that Re($p$) > 0. 
This solution must satisfy Im($p$) = 0 because otherwise 
quantities such as $\langle x^2 \rangle$ would take negative values. But it is impossible 
for $p$ to always take a real positive value for the following reason. 
As $\delta  \rightarrow 0^+$, the term proportional to $\delta^4$ becomes unimportant,
and note that $H(p)/J(p)$ has a lower positive bound as it is a ratio of positive polynomials and tends to infinity as $p \rightarrow \infty$. 
We have thus reached a contradiction which means that the system is stable for 
sufficiently small $\delta$. 

Having established the first part of the argument, we now want to obtain 
the value of $\delta$ at which the system becomes unstable. Any onset of 
instability must to go through $p=0$ [again, Im($p$) must vanish]. We set $p=0$ in 
(\ref{determinantal}) and find that 
there is only one solution. Reassuringly, this solution is exactly the one given
by (\ref{deltacrit}). 
The system is thus stable for $\delta < \delta_c$.


\bsp	
\label{lastpage}
\end{document}